\documentclass[%
aip,
amsmath,amssymb,
reprint,%
]{revtex4-1}

\usepackage{graphicx}% Include figure files
\usepackage{dcolumn}% Align table columns on decimal point
\usepackage{bm}% bold math
\usepackage{color,framed}
\usepackage{textcomp}
\usepackage{soul}
\soulregister\cite7 % 閽堝\cite鍛戒护
\soulregister\ref7 % 閽堝\ref鍛戒护
\soulregister\pageref7 % 閽堝\pageref鍛戒护
\usepackage[utf8]{inputenc}
\usepackage[T1]{fontenc}
\usepackage{mathptmx}
\begin{document} 
 
%\preprint{ }
 
\title[ ]{Direct temperature determination of {{a}} sympathetically cooled large $^{113}$Cd$^{+}$ ion crystal for {{a}} microwave clock}
 
 \author{Y. N. Zuo}
 \author{J. Z. Han}%
 \affiliation{Department of Physics, Tsinghua University, Beijing 100084, China
 %\\This line break forced with \textbackslash\textbackslash
}
 \affiliation{%
 	{{State Key Laboratory of Precision Measurement Technology and Instruments, Tsinghua University, Beijing 100084, China}}
 }

 \author{J. W. Zhang}
 \email{zhangjw@mail.tsinghua.edu.cn}
\affiliation{%
	{{State Key Laboratory of Precision Measurement Technology and Instruments, Tsinghua University, Beijing 100084, China}}
}
 \affiliation{%
 Department of Precision Instruments, Tsinghua University, Beijing 100084, China
}

 \author{L. J. Wang}
 \email{lwan@mail.tsinghua.edu.cn}
 \affiliation{Department of Physics, Tsinghua University, Beijing 100084, China
 	%\\This line break forced with \textbackslash\textbackslash
 }
 \affiliation{%
 	{{State Key Laboratory of Precision Measurement Technology and Instruments, Tsinghua University, Beijing 100084, China}}
}
 \affiliation{%
 Department of Precision Instruments, Tsinghua University, Beijing 100084, China
}

 \date{\today}% It is always \today, today,
 % but any date may be explicitly specified
 
\begin{abstract}
This paper reports direct temperature determination of sympathetically cooled $^{113}$Cd$^{+}$ ions with laser-cooled $^{24}$Mg$^{+}$ in a linear Paul trap. The sympathetically cooled ion species distribute in the outer shell of the large ensembles{{, which contain}} up to $3.3 \times 10^{5}$ ions. With optimized parameters, the minimum temperature of the sympathetically cooled $^{113}$Cd$^{+}$ ions {{was}} measured to be {on the order of 10 mK}. These results {are promising for performance of} microwave atomic clocks. {{The second-order Doppler frequency shift is two orders of magnitudes lower (from {$1.88 \times 10 ^{-14} $ to $6.26 \times 10^{-16} $})}} and the Dick effect is suppressed. %{\cite{Fordell:19}}
\end{abstract}

\maketitle
 
% \section{Introduction}
 
Charged particles in ion traps {{are more strongly}} bound {than} neutral atoms in magneto-optical traps. With lasers, trapped ions can be cooled translationally {{by Doppler cooling only to the millikelvin level, and need further cooling down to the vibrational ground state {on the microkelvin scale}.\cite{Huntemann2016,Chen2017}}} {Hence, ion trapping is} used widely in precision measurements, such as frequency metrology,\cite{Huntemann2016} mass spectrometry,\cite{Deb2015} precision spectroscopy,\cite{Biesheuvel2016} quantum information processing,\cite{Zhang2017} measuring physical constants,\cite{Huntemann2014} and chemical physics.\cite{Willitsch2012} Frequency standards based on trapped ions play an important role in frequency metrology. {{In particular,}} microwave frequency standards based on ions {{have good potential for {use in} compact atomic clocks,\cite{mulholland2018portable,schwindt2015miniature,gulati2018miniatured}}} deep-space navigation, ultra-stable timekeeping, and space-borne clocks.\cite{Ely2018} 

The microwave clock based on laser-cooled $^{113}$Cd$^{+}$ {{had a}} short-term frequency instability of $6.1 \times 10^{-13} / \sqrt{\tau}$ and {{a}} frequency uncertainty of $6.6 \times 10^{-14}$.\cite{Miao2015} Nevertheless, it {{has not reached}} the performance anticipated {{by Zhang \emph{et al.}}}\cite{Zhang2015} This kind of microwave frequency standard, based on laser-cooled ions, {{generally suffers from the Dick effect {owing} to the dead time for the operation and the second-order Doppler frequency shift (SODFS) from ion temperature {rising} {during the clock interrogated}.\cite{JohnDick1990}}} Sympathetic cooling (SC) is promising for overcoming both of these {limitations}. SC can cool the translational motion of the target atomic or molecular ions via mutual Coulomb interaction. {{It was first realized for isotopic ions in a Penning trap.\cite{drullinger1980}}} It has been demonstrated with laser-cooled alkaline-earth {and} alkaline-like-earth metal ions, such as Be$^{+}$, Mg$^{+}$, Ca$^{+}$, Ba$^{+}$, and Yb$^{+}$. The Doppler limit (${<}1$~{{mK}}) {has been} reached for small crystals {consisting} of about 10 ions, {as} verified by spatial thermometry.\cite{knunz2012} By applying SC, the temperature of Cd$^+$ ions can be kept low during the whole interrogation sequence, and no additional cooling {is} needed. Because the {ion temperature} is low, the SODFS and uncertainty {in} SODFS due to ion motion can be significantly reduced. Moreover, the Dick effect can be suppressed {because the dead time is shorter than for a directly cooled Cd$^+$ clock.}

{We have proposed using} $^{24}$Mg$^{+}$ as the coolant to sympathetically cool {{a}} large $^{113}$Cd$^{+}$ cloud.\cite{zhang2015a} {In contrast with} other ions, $^{24}$Mg$^{+}$ {only needs one laser to cool it}. The large mass difference between {the} Cd$^{+}$ and Mg$^{+}$ ions {(mass ratio = 4.71)} results in a significant spatial separation between these two kinds of ion crystal, which is helpful {{for reducing}} the AC Stark shift caused by laser {cooling the $^{24}$Mg$^{+}$ ions}. To estimate the SODFS, the exact temperature of $^{113}$Cd$^{+}$ has to be obtained. One can measure the Doppler broadening of a spectral transition to determine the ion temperature.\cite{hornekaer2000,van1999,hasegawa2001} However, {it is difficult to measure the ion temperature directly without suitable direct-cooling lasers}. {{An indirect method for measuring the ion temperature is to compare}} CCD photographs of the ions to statistical images generated by {a} molecular dynamics (MD) simulation.\cite{offenberg2008} Nevertheless, the accuracy of this method is limited by the hypotheses {{used in the}} simulation models and the computational accuracy. There are several reports of direct measurements of SC ion temperature.  For example, Wineland {\emph{et al.}} measured the temperature of $^{199}$Hg$^{+}$ sympathetically cooled by $^{9}$Be$^{+}$ {{in Penning traps}} to {{0.4--1.8~K}},\cite{larson1986} and Imajo {\emph{et al.}\ measured the temperature of SC Cd$^{+}$ cooled to 1~K using isotopic ions, {{also in Penning traps}}.\cite{imajo1996} {The direct temperature of the inner Mg$^{+}$ core of Ca$^{+}$--Mg$^{+}$ bi-crystals was measured as 40~mK in a Paul trap.\cite{hornekaer2000}} Reported experiments {{show}} that it {{is}} very challenging to cool target ions {{sympathetically using}} coolant ions with large mass differences because of the limited cooling efficiency.\cite{imajo1997}
	
In this letter, we report the {direct temperature measurement} of $^{113}$Cd$^{+}$ sympathetically cooled by $^{24}$Mg$^{+}$. {{This can be used in}} a highly accurate microwave atomic clock. {The temperature of the outer shell of a sympathetically cooled ion crystal was directly measured to be on the sub-kelvin scale with as many as $3.3 \times 10^{5}$ ions and a large mass ratio in a Paul trap. The SODFS {{was}} reduced to two orders of magnitude lower than that of laser-cooled ion clouds.\cite{Miao2015}} {{Moreover,}} the Dick effect {{was}} also reduced owing to a shorter dead time, {because} no additional cooling {was} needed. Hence, {this makes it possible to} build high-performance microwave atomic clocks based on SC ions. This result can also be treated as an upper limit for the translational temperature of outer-shell SC ions {{when refining}} MD simulation models.

% \section{EXPERIMENT}
 
The experimental setup {was} an enhanced version of {that} described in previous reports.\cite{Miao2015,Zhang2015} {{We used}} a linear quadrupole Paul trap made of oxygen-free copper. {{It had}} four parallel cylindrical electrodes {each} with a radius of $r_{e}=7.1$~mm. Each electrode {was} divided into three segments, {{of lengths}} 20, 40, and 20~mm. The minimum distance from the trap center axis to the {surfaces} of the electrodes {was} $r_{0}=6.2$~mm. The ratio of $r_{e}$ {to} $r_{0}$ {was} close to the optimized value for an ideal quadrupole field.\cite{denison1971} A radio frequency (RF) voltage $V_{\mathrm{RF}}$ {{was}} applied to one pair of diagonal electrodes, and the other pair {{was}} grounded. The driving frequency {{was}} {$\Omega = 2 \pi \times 2.02$~MHz}. A static voltage $U_{\mathrm{End}}$ was added to the end-cap electrodes, which {could} be adjusted between 0 and 120~V to manipulate the shapes of the Coulomb crystals. The trapping parameters \cite{paul1990} of the well-known Mathieu equation {{are}} $q_{r}^{\mathrm{Mg}}=0.03$ and $q_{r}^{\mathrm {Cd}}=0.15$, and the trap depths are $d^\mathrm{Mg}=2.11 $~eV and $d^\mathrm{Cd}=0.45 $~eV. 
 
To manipulate the Cd$^{+}$ and Mg$^{+}$ at the same time, two frequency-quadrupled lasers with wavelengths of 214 and 280~nm {counter-propagated} along the trap axis. The diameters of the two laser beams {were} both 1~mm, and {{the intensities}} {were} 6~mW/mm$^{2}$ at 280~nm and 1~mW/mm$^{2}$ at 214~nm. The frequencies {were} stabilized with high-precision wavemeters {at} the transition lines of 3s$^{2}$S$_{1/2}$ $\leftrightarrow$ 3p$^{2}$P$_{1/2}$ of $^{24}$Mg$^{+}$ and 5s$^{2}$S$_{1/2}$ $\leftrightarrow$ 5p$^{2}$P$_{3/2}$ of $^{113}$Cd$^{+}$. The natural {width} the D$_{1}$ line for $^{24}$Mg$^{+}$ is $2\pi \times 42.7$~MHz, which gives a minimum Doppler temperature of 1.0~mK,\cite{martin1980} while the natural {width} of the D$_{2}$ line for $^{113}$Cd$^{+}$ is $2 \pi \times 60.1$~MHz, corresponding to 1.4~mK.\cite{moehring2006} {{The ions {were} imaged}} by an electron-multiplying charge-coupled device (EMCCD) camera. The camera system, {{which {was}}} mounted on a precision motion stage, {consisted} of a lens system, tunable pinhole, and ultraviolet (UV) filters. The magnification {{factor}} of the lens {could} be adjusted from 4 to 8. The exposure time {was} set to 0.3~{{seconds}} for one shot. In addition, {we used} a photo-multiplier tube with optional filters at the opposite side of the trap axis to {{monitor the fluorescence intensity}}.

To load the ions, neutral atoms {were} evaporated from pure metal in different ovens. The $^{24}$Mg isotope {was} at the natural abundance, while $^{113}$Cd {was} isotopically enriched. Magnesium atoms {were} ionized by electron bombardment, and cadmium atoms {were} ionized by resonance-enhanced two-photon ionization using {{another frequency-quadrupled laser}} at 228 nm. Photo-ionization for $^{113}$Cd$^{+}$ avoids disturbing and heating the loaded Mg$^{+}$.{{\cite{hornekaer2000}}} Photo-ionization {further} has the advantages of a controllable high loading efficiency and isotopic selection{{, and it does not produce a stray electric field like that from the residual charge of an electron gun.}} The Mg$^{+}$ ions were loaded {{first}}, and {the} Cd$^{+}$ ions were {then} loaded. {{The ionization energy {was} set to}} avoid {production} of $^{24}$Mg$^{2+}$, whose ionization energy is 22.7~eV.\cite{hornekaer2000,drewsen2003} While loading Mg$ ^{+} $, $V_{\mathrm{RF}}$ {was} set to a higher value than {for} normal detection to {{produce}} a deeper potential depth to load more ions. The compensation voltages applied to the electrodes have to be tuned carefully after loading to obtain symmetric ion crystals. During the loading, many of the $^{24}$Mg$^{+}$ ions {are} converted to $^{24}$MgH$^{+}$ dark molecular ions through chemical reactions with residual H$_{2}$ molecules in the background gas,\cite{molhave2000} {{which reduces the number of ions and decreases SC efficiency}}. {{In contrast, the production rate of CdH$^{+}$ is}} quite low.
 
\begin{figure}[tbp]
 \centering
 \includegraphics[width=0.9\linewidth]{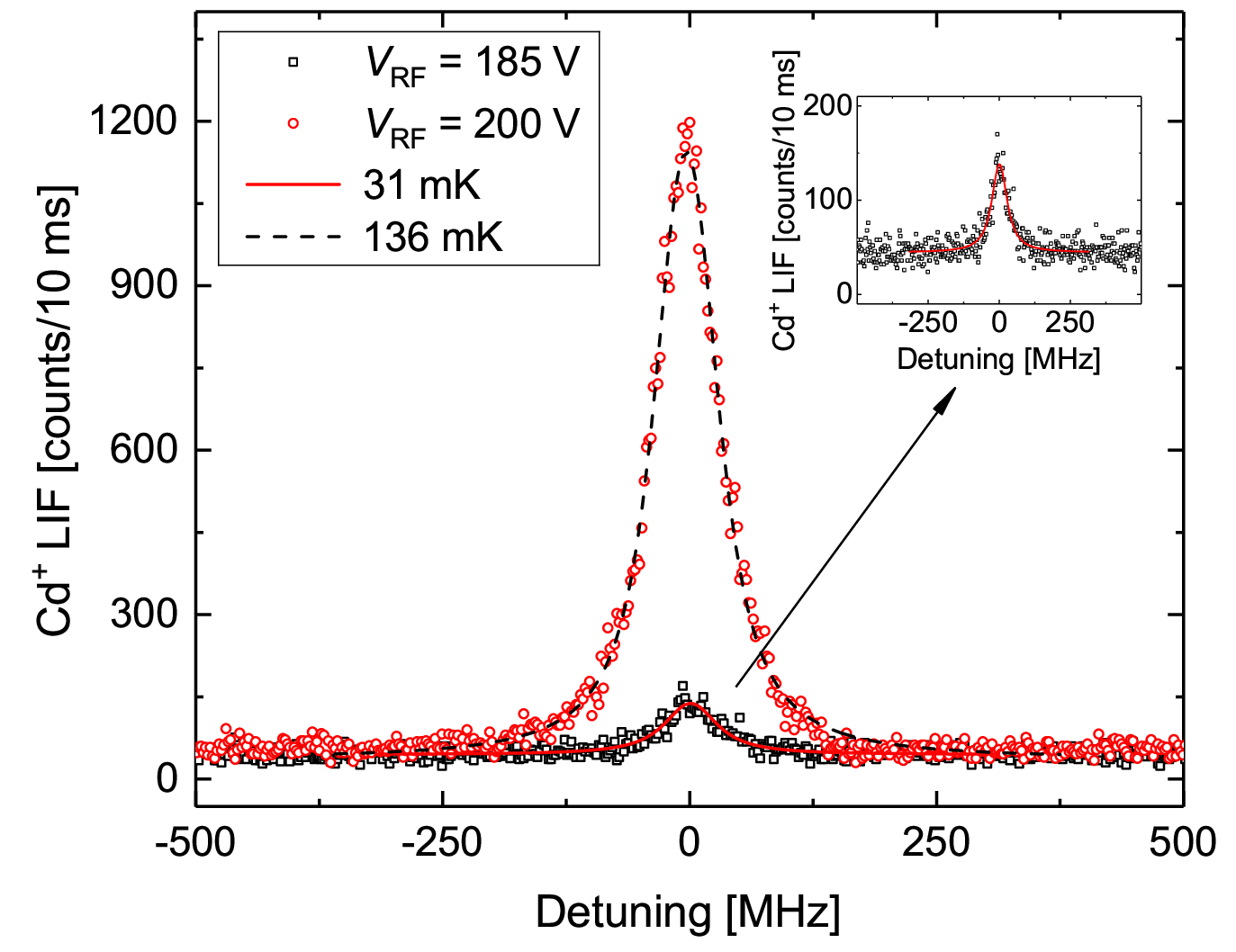}
 \label{fig:uselif}
 \caption{Fluorescence spectrum of SC $^{113}$Cd$^{+}$. Experimental data are marked by squares and circles for $V_{\mathrm{RF}} = 185$~V and 200~V, respectively. The curves are fitted {with} a Voigt function {having} a Lorentz linewidth of 60.13~MHz, which is the natural width of the D$_{2}$ transition of $^{113}$Cd$^{+}$. The temperatures obtained for $^{113}$Cd$^{+}$ are 31~mK and 136~mK, respectively. The inset shows the fitting of the 31~mK result.}
 %The frequency of probing laser scans past the 5s$^{2}$S$_{1/2}$ F=1 to 5p$^{2}$P$_{3/2}$ F=2 transition of $^{113}$Cd$^{+}$ ions with a range of 1 ${GHz}$. The laser power is set low enough to avoid cooling or heating effects. 
\end{figure}
 
To {determine} the temperature of {{SC}} Cd$^{+}$, {we measured} the Doppler broadening of the transition {from} 5s$^{2}$S$_{1/2}$ $F=1$ to 5p$^{2}$P$_{3/2}$ $F=2$. During this measurement, the {{intensity}} of the 214~nm laser {was} set to 20~\textmu W/mm$^{2}$ to avoid cooling or heating the ions. The frequency of 214~nm laser {was} scanned around the resonant frequency with a range of {{1~GHz}}. The measured line profiles {were} fitted {with} a Voigt function. The Lorentz linewidth of the Voigt function {was} set {as} the natural linewidth of the D$_{2}$ transition of $^{113}$Cd$^{+}$, which is 60.13~MHz ($2.647 \pm 0.010$~ns).\cite{moehring2006} The Gaussian linewidth {{from the fitting}} represents the velocity distribution of {the} SC Cd$^{+}$ ions. Thus, the temperature can be calculated {via\cite{riehle2006frequency,warrington2002temperature,drewsen2002ion}}
%\begin{equation} 
%\Delta v_{\mathrm{G}}=\frac{v_{0}}{c} \sqrt{ \frac{8 \ln 2 \cdot k_{\mathrm{B}} T}{M}},
%\end{equation}
\begin{equation} 
 T=\frac{M c^{2}}{8 \ln 2 \cdot k_{\mathrm{B}} }\left(\frac{\Delta v_{G}}{v_{0}}\right)^{2},
\end{equation}
 where ${M}$ is the mass of $^{113}$Cd$^{+}$, ${c}$ is the speed of light, ${k}$ is {the} Boltzmann constant, ${\Delta v_{G}}$ is the Gaussian linewidth, and ${v_{0}}$ is the resonant frequency of the D$_{2}$ transition of Cd$^{+}$. {Figure 1 shows {{a}}} typical Voigt fitting of the laser-induced fluorescence (LIF) spectra. {The LIF intensity is proportional to the number of resonant ions, and the ion temperature} depends on $V_{\mathrm{RF}}$ and the number of ions. 
 As shown in Fig.~1, the temperature is 136~mK when $V_{\mathrm{RF}}=200$~V and 31~mK when $V_{\mathrm{RF}}=185$~V. These results indicate that the {ion temperature} is sensitive to the variation of $V_{\mathrm{RF}}$ for SC when there is a large mass difference between the two ion species. 
 
% \section{RESULTS AND DISCUSSION}
% \begin{figure}
% \includegraphics{1-sequcen-1}% Here is how to import EPS art
% \caption{\label{fig:epsart} Schematic of a typical temperature determination sequence. (1) Doppler cooling of Mg+. (2) tuning compensated voltages applied to the electrodes to decrease RF heating. (3) loading and cooling for Cd$^{+}$. (4) measurement.}
% \end{figure}

Ions with different charge-to-mass ratios {{have}} different displacements {owing} to the mass dependence of the pseudo-potential. In large {{multi-component}} Coulomb crystals, such displacements can result in a complete {radial} separation.{\cite{offenberg2008}} {{Moreover,}} the different radiation pressure{s} of the cooling lasers segregates {{the}} ions axially. {{In Fig.~2, only the axial separation is clearly shown. A clear radial separation is shown in Fig.~3, which is for fewer ions than in Fig.~2.}}

{We had to photograph the entire multi-component ion crystal to obtain its configuration and dimensions.} {The crystal had a large aspect ratio owing to the relatively weak axial confinement field.} {{Because}} the diameter of the 214.5~nm laser beam {was} 1~mm, the diameters of {the} Cd$^{+}$ ion clouds {were} larger than those of the laser beams. Thus, only part of the Cd$^{+}$ {could} be imaged {{in a single exposure}} {owing} to the limited EMCCD image area. {{Composite photographs of the entire ion cloud can be}} obtained by stitching together 12 EMCCD photos, as shown in Fig.~2. {{Moreover,}} {because of the lens} chromaticity at 214.5~nm and 280~nm, the focus positions are different for Mg$^{+}$ and Cd$^{+}$. This {{difference was}} taken account in combining the photographs. 
 
\begin{figure}[tbp]
 \centering
 \includegraphics[width=0.9\linewidth]{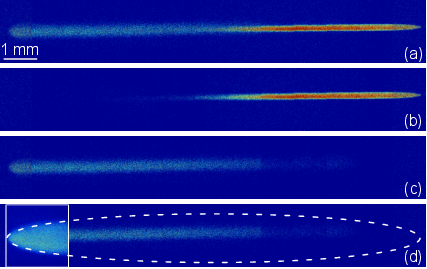}
 \label{fig:3-use-full-4}
 \caption{{Combination} photographs of Cd$^{+}$ and Mg$^{+}$ captured by the EMCCD with UV filters and a 0.3~{{s}} exposure time. The total number of ions is $N_\text{tot}=3.3 \times 10^{5}$. Each picture {{is a composite of 12 separate}} EMCCD photographs. {Panel (a) is a} composite of (b) and (c), with {$V_{\mathrm{RF}} = 280$~V} and $U_{\mathrm{End}} = 10$~V, where (b) and (c) show Mg$^{+}$ and Cd$^{+}$, {{respectively}}. These two kinds of ions are pushed to different sides {{of the cloud}} {owing to the unidirectional} laser incidence. The {{small}} difference in the magnification factor due to the chromaticity is taken into account. The {{left}} end of (d), marked by the solid {box}, is obtained by overlaying 10 photographs taken for different positions of the laser beam for Cd$^{+}$. The profile of the entire ion cloud is marked by the {dashed} line.}
\end{figure}

%, and ${r_{0}}$ is the minimum distance from the trap center to the electrodes鈥?surface

The numbers of the two visible atomic ions were {{estimated}} {using} the size of the ion crystal {{in the}} EMCCD images and the {{calculated ion}} number density. Using the {zero-temperature charged-liquid} model,{\cite{hornekaer2000,hornekaer2001}} the number density ${n_{0, i}}$ under the pseudo-potential approximation is 
 \begin{equation} 
 n_{0, i}(r, z)=n_{i}=\frac{\varepsilon_{0} V_\text{RF}^{2}}{M_{i} \Omega^{2} r_{0}^{4}},
 \label{equ:density}
 \end{equation}
where {${i}$ indicates the ion species,} ${\varepsilon_{0}}$ is the permittivity of a vacuum, $V_{\mathrm{RF}}$ is the voltage applied to the electrodes, ${M_{i}}$ is the {ion} mass, ${k}$ is {the} Boltzmann's constant, and ${\Omega}$ is the trap driving frequency. {{Thus, the number densities for Mg$^{+}$ and Cd$^{+}$ are}} {{$n_{0, \mathrm{Mg}} = 7.32 \times 10^{13}$~m$^{-3}$ and $n_{0,\mathrm{Cd}} = 1.55\times 10^{13}$~m$^{-3}$,} respectively, with $V_{\mathrm{RF}}=280$~V. 
 
To obtain the total number of {ions in the multi-component crystal}, we have to estimate the total volume of the ions. The Mg$^{+}$ crystal in the core {is} approximated as a cylinder and the Cd$^{+}$ crystal as an ellipsoid shell. From Fig.~2, the three radii of the ellipsoid shell are 6.45~mm, 0.89~mm, and 0.89~mm. Moreover, the length of the inner cylinder is {{6.45~mm}} and {{its}} radius is 0.14~mm. The gap between the inner surface of the ellipsoid shell and the outer surface of the inner cylinder is proportional to the root of the mass ratio and the inner radius,\cite{hornekaer2000,o1981} {{which we calculated}} to be {approximately} 0.16~mm. Thus, the number of ions {is} estimated to be about $N_\text{tot}=3.3 \times 10^{5}$ with {{$N_\text{tot,Mg}=2.9 \times 10^{4}$ and $N_\text{tot,Cd} = 3.0 \times 10^{5}$.}} It {is} difficult to determine accurately the number of each {{type}} of ion species under the present conditions {owing} to the large total volume of the ions. {There were some dark ions in the crystal whose species were not identified. Because the dark ions can be confined in the trap, their masses should be close to those of Mg$^{+}$ and Cd$^{+}$. According to the literature,\cite{molhave2000} the dark ions could be MgH$^{+}$. We have ignored the dark ions in estimating the total number of ions.}
 
 \begin{figure}[tbp]
 \centering
 \includegraphics[width=0.9\linewidth]{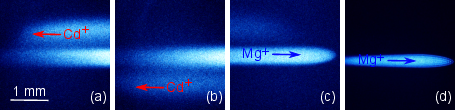}
 \label{fig:4-use-sigle-4}
 \caption{EMCCD images of a {{multi-component}} ion crystal for {a 3~s exposure time}. {Panels} (a), (b), and (c) {{were}} taken simultaneously without UV filters. The red arrow indicates the propagating direction of the 214~nm laser, and the blue one {is that for} the 280~nm laser. {Panels (d) is a clear} EMCCD image of a $ ^{24} $Mg$^{+}$ ion crystal taken with {a} 280~nm filter.}
 \end{figure}

To {{get insight into}} the separation of the two ion species, {Figures}~3{{(a)--(c)} show the complete spatial separation between $^{24}$Mg$^{+}$ and $^{113}$Cd$^{+}$. {{These images}} were taken without {{optical}} filters. {{In}} Figs.~3(a) and 3(b), the 214.5~nm laser {{was}} tuned to brighten the bottom and upper {{parts}} of Cd$^{+}$, respectively. In Fig.~3(c), the profile of Mg$^{+}$ in the core is clear. {These were taken with an integration time of 3~s}. {{In the same photograph,}} the profile of Cd$^{+}$ is fuzzy {owing} to chromatic aberration. Figure~3(d) shows a small Mg$^{+}$ ion crystal that has crystallized well. {{{This image} was taken using a 280~nm optical filter.}}
 
Although the first-order Doppler effect is suppressed {{{owing} to}} the Lamb--Dicke criterion, the SODFS is not negligible for microwave clocks {{based on ions}}. {{Lowering}} the temperature helps to reduce the SODFS. {{Hence, we measured the temperature of the ions for different values of $V_{\mathrm{RF}}$ and $U_{\mathrm{End}}$. The results are shown in Figs.~4 and~5, respectively.}} {Figure 4 shows} the temperature {dependence} of outer-shell SC ions on the trapping voltage $V_{\mathrm{RF}}$. {{All the data for this figure were measured for the same multi-component crystal in a single loading.}} Each point is the average of three measurements to ensure {{that equilibrium has been reached.}} {Each} error bars {is} the maximum difference between the three measurements {in that set}. After preparing a large multi-component crystal, {{{we tuned} the trapping voltages}} in a small range. {{For each value of $U_{\mathrm{End}}$, we measured the ions temperature by}} changing $V_{\mathrm{RF}}$ in two opposite {directions}. {{The blue squares {represent stepping up} $V_{\mathrm{RF}}$ from a low level to a high level, and the red triangles {correspond to stepping down} $V_{\mathrm{RF}}$.}} The maximum measured temperature was about 200~mK, and the minimum temperature was {on the order of 10 mK}.
 
   \begin{figure}[tbp]
  	\centering
  	\includegraphics[width=0.9\linewidth]{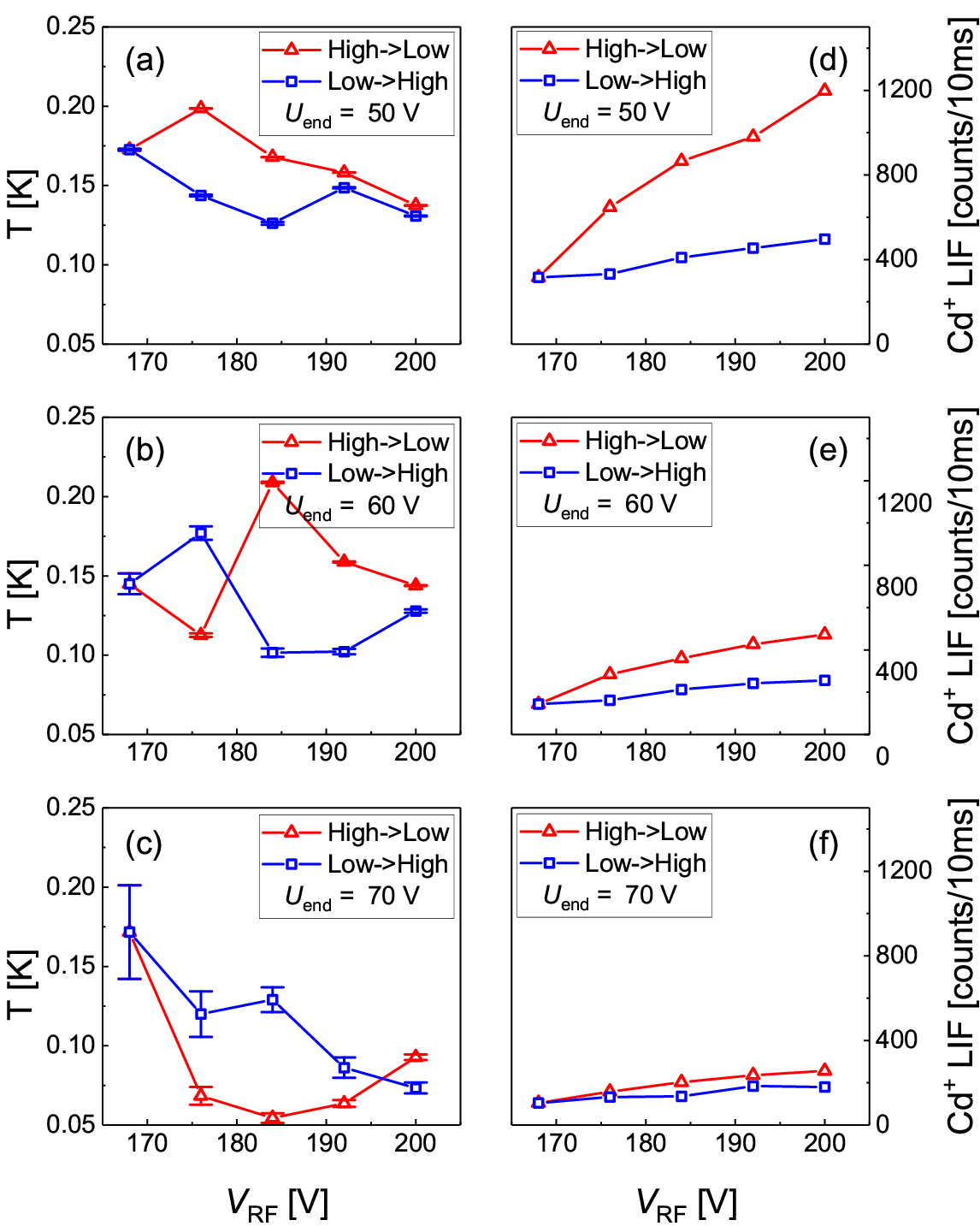}
  	\label{fig:5-rf-lif}% Here is how to import EPS art
  	\caption{{{The trapping RF voltage $V_{\mathrm{RF}}$ affects the SC efficiency. The graphs show the temperature of}} Cd$^{+}$ while the RF voltage {{are scanned for}} different end-cap voltages. {Panels (a)-(c)} show the variation of temperature with RF voltage, while {(d)-(f)} show the LIF signal intensity for these detections.}
  \end{figure}

In {Figures~4(a)-(c)}, we can see that there is no simple correlation between the ion temperature and $V_{\mathrm{RF}}$. For example, the ion temperature could descend, ascend, and then descend again as $V_{\mathrm{RF}} $ increases. Thus, the ion temperature may vary with $V_{\mathrm{RF}} $ in different ways. The dependence of the ion temperature on $V_{\mathrm{RF}} $ strongly depends on the RF heating effect and the number of ions. The RF heating effect is due to ion--ion interactions during which micromotion energy is transferred to secular energy. Conversely, the RF heating depends on the ion temperature.\cite{ryjkov2005simulations} For example, in mono-component Mg$ ^{+} $ crystals, the RF heating rate {is} relatively low when the ion temperature {is} below 0.5~K, but {increases} rapidly when the ion crystal {undergoes} a phase transition.\cite{ryjkov2005simulations, zhao2006fluorescence} It is unclear how the RF heating {affects} the temperature of a large multi-component plasma (with up to $10^{5} $ ions).\cite{hornekaer2002formation} However, it is clear that the ion--ion Coulomb collision rate decreases both at high and low temperatures.\cite{ryjkov2005simulations} When ions temperature decreases, the amount of heat exchanged between the two components decrease{s}, which decrease{s} the efficiency of both the RF heating and SC. However, the rate{s at which the RF heating and SC efficiencies fall} could be different. Thus, the irregular changes in the ion temperature in Fig.~4 might be caused by competition between RF heating and SC, or a structural phase transition.

In addition, the number of ions decreases with time, which results in a collective decrease {in} signal intensity. {This is} even for the same probe laser intensity, as shown in Figs.~4(d){-}(f). The ion density depends on the RF voltage for the same $U_{\mathrm{End}}$ {value}. Thus, the fluorescence counts increase as $V_{\mathrm{RF}}$ increases. However, a lower number of ions does not correspond to a lower temperature {owning} to the competition between RF heating and SC, as shown in Figs.~4(c) and 4(f). More importantly, the nonlinear resonance effects in a linear Paul trap with circular (rather than hyperbolic) electrodes {are} not negligible,\cite{takai2007nonlinear} like the nonlinear resonance due to Coulomb interactions among ions.

 \begin{figure}[tbp]
 \centering
 \includegraphics[width=0.9\linewidth]{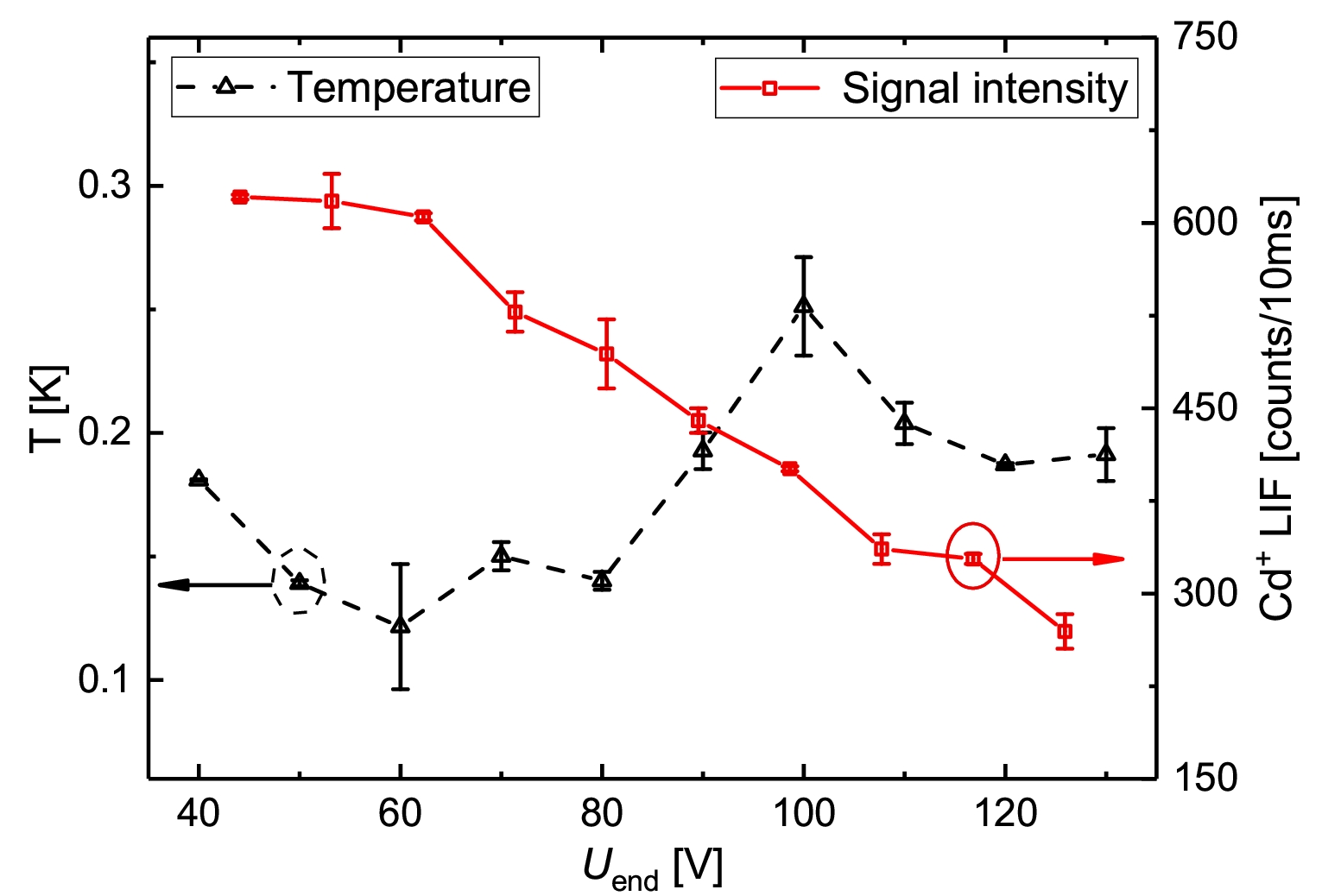}
 \label{fig:6-end-lif}
 \caption{End-cap voltage $U_{\mathrm{End}}$ affects the SC efficiency. The RF voltage $V_{\mathrm{RF}}$ {is} fixed at 200~V.}
 \end{figure}
 
The {ion temperature dependence} on $U_{\mathrm{End}}$ is shown in Fig.~5. The black triangles represent temperatures, and the red squares represent fluorescence count {for varying $U_{\mathrm{End}}$}. Because of the Coulomb interaction with SC and the unbalanced radiation pressure of the cooling laser, the ion temperature decreases {and} then increases as $U_{\mathrm{End}} $ increases. The minimum temperature {is} at 60~V. A similar phenomenon {has been} observed for a small {{mono-component}} ion crystal (with 34 Ca$^{+}$ ions).\cite{okada2010} The spatial radius of each Ca$^{+}$ ion also reaches a minimum at a relatively low voltage. In Fig.~5, the number of trapped ions decreases as $U_{\mathrm{End}} $ increases, leading to a decrease in the signal intensity. Hence, the temperature for a relatively high $U_{\mathrm{End}}$ (${>} 100$~V) again decreases.

{{An}} accurate evaluation of the SODFS would require MD simulations from which the ion velocity distribution could be extracted. {{However,}} it can be roughly estimated using simple models for microwave ion clocks. The total SODFS due to both the secular motion and the micromotion of trapped ions is estimated as \cite{prestage1999higher}
\begin{equation}
 \frac{\Delta f}{f_{0}}
 = \left\langle-\frac{E_\text{Kinetic}}{M c^{2}}\right\rangle
 = -\frac{3}{2} \frac{k T}{M c^{2}}\left[1+\frac{2}{3}\left(N_{d}^{K}\right)\right],
\end{equation}
where ${f_{0}}$ is the central frequency, and ${N_{d}^{K}}$ is {the SODFS coefficient} due to the micromotion averaged across the whole ion cloud. {The temperature of $^{113} $Cd$ ^{+} $ decreases from $ 6 \pm 1 $~K\cite{Miao2015} to $0.20 \pm 0.05 $~K, which reduces the SODFS from $1.88 \times 10 ^{-14} $ to $6.26 \times 10^{-16} $. Moreover, the uncertainty for this shift also decreases from $0.31 \times 10^{-14}$ to $1.57 \times 10^{-16} $.}	

For a cadmium ion microwave clock, dead time is unavoidable during interrogation, {hence the Dick effect degrades the short-term frequency stability}. {{With}} the SC method, the dead time can be reduced {because} no additional cooling step is needed. Thus, the Dick-effect-limited Allan deviation falls from $4 \times 10^{-13} / \sqrt{\tau}$ to $2 \times 10^{-13} / \sqrt{\tau}$ if the same local oscillator and microwave synthesizer are used{{, as in Zhang \emph{et al}}}.\cite{Zhang2015} In addition, illumination by a 280~nm laser could cause an AC Stark shift of the hyperfine clock transition. Fortunately, this can be controlled to {be as low as the} fluorescence by making the beam thin enough.\cite{zuo2016progress} 
 
% \section{Conclusions}

In summary, we produced a large SC ion crystal with up to $3.3 \times 10^{5}$ ions and a large mass ratio of 4.71 {{in a Paul trap}}. {{And we measured its}} temperature. The {{current}} mass ratio is based on the obtainable laser wavelengths, without optimization. We obtained {a} lower temperature for the outer shell even with {{a}} larger mass ratio. However, there is no clear definition of SC efficiency. It can intuitively be qualified by the mass ratio, the acquired temperature, and the total number of SC ions. This is the lowest {{directly measured temperature}} for such a large number of outer-shell SC ions with such a high mass ratio. These results {confirm} the translational temperature obtained by MD simulations. {{The SODFS of $ ^{113} $Cd$ ^{+} $ was reduced by two orders of magnitude and the Dick-effect-limited Allan deviation was halved under the same conditions as in our previous research.\cite{Zhang2015}} These results may be important in building a SC Cd$^{+}$ microwave frequency standard and useful for research on the SC of large ion crystals. 
 
\begin{acknowledgments}
 This work {was} supported by the National Key R\&D Program of China (2016YFA0302101) and the Initiative Program of State Key Laboratory of Precision Measurement Technology and Instruments.
\end{acknowledgments}

\nocite{*}
% \bibliography{aipsamp}% Produces the bibliography via BibTeX.
% \bibliography{bibfile}

\providecommand{\noopsort}[1]{}\providecommand{\singleletter}[1]{#1}%

\end{document}